\documentclass[runningheads]{llncs}

\usepackage[T1]{fontenc}
\usepackage{multirow}
\usepackage{multicol}
\usepackage{siunitx}
\usepackage[table,dvipsnames]{xcolor}
\usepackage{graphicx}
\usepackage{cite}
\usepackage{mdframed}
\usepackage{url}
\usepackage{hyperref}
\usepackage{hhline}
\usepackage{amsmath}
\usepackage{cleveref}
\usepackage{subfigure}
\usepackage{bbding}

\begin{document}
\title{Post-Training Denoising of User Profiles with LLMs in Collaborative Filtering Recommendation\thanks{This preprint has not undergone peer review (when applicable) or any post-submission improvements or corrections.}}
\titlerunning{Post-Training Denoising of User Profiles with LLMs in CF Recommendations}
\author{
Ervin Dervishaj\Envelope \inst{1} \and
Maria Maistro \inst{1} \and
Tuukka Ruotsalo \inst{1,2} \and
Christina Lioma \inst{1}
}

\authorrunning{Dervishaj et al.}

\institute{University of Copenhagen \and LUT University \\
\email{\{erde,mm,tr,c.lioma\}@di.ku.dk}}

\maketitle

\begin{abstract}
Implicit feedback -- the main data source for training Recommender Systems (RSs) -- is inherently noisy and has been shown to negatively affect recommendation effectiveness. Denoising has been proposed as a method for removing noisy implicit feedback and improving recommendations. Prior work has focused on \emph{in-training} denoising, however this requires additional data, changes to the model architecture and training procedure or fine-tuning, all of which can be costly and data hungry.
In this work, we focus on \emph{post-training} denoising. Different from in-training denoising, post-training denoising does not involve changing the architecture of the model nor its training procedure, and does not require additional data. Specifically, we present a method for post-training denoising user profiles using Large Language Models (LLMs) for Collaborative Filtering (CF) recommendations. Our approach prompts LLMs with (i) a user profile (user interactions), (ii) a candidate item, and (iii) its rank as given by the CF recommender, and asks the LLM to remove items from the user profile to improve the rank of the candidate item. Experiments with a state-of-the-art CF recommender and 4 open and closed source LLMs in 3 datasets show that our denoising yields improvements up to 13\% in effectiveness over the original user profiles. Our code is available at \url{https://github.com/edervishaj/denoising-user-profiles-LLM}.

\end{abstract}

\section{Introduction and Related Work} \label{sec:intro}
Recommender Systems (RSs) are integral to delivering relevant and personalized user experiences. As the number of items on online platforms continues to grow, users tend to engage with only a small subset of 
items and 
give feedback on an even smaller fraction of those interactions~\cite{hu2008collaborative}. Such data sparsity is 
a key obstacle for learning user preferences from interactions. For this reason, implicit user feedback has been typically used to build user profiles and, in turn, develop more effective recommendation models. 
Although implicit feedback is generally more abundant than explicit feedback, it is inherently noisy \cite{hu2008collaborative,joachims2017accurately}, due to factors, such as fluctuations in user mood, time-dependent behaviors, account sharing, for instance. 
Such factors introduce inconsistencies that obscure users’ true preferences \cite{joachims2002optimizing,joachims2007evaluating}, and can lead to significant drops in recommendation quality\cite{wang2021denoising}.

Denoising has been proposed as a method for extracting cleaner user preference signals from implicit feedback~\cite{wang2021denoising}. However, removing noisy implicit feedback is non-trivial due to the lack of ground truth labels, i.e., labels that identify an interaction as a noisy one. 
Prior work on denoising uses auxiliary information regarding the user-item interaction, like user dwell time~\cite{yi2014beyond,kim2014modeling} or mid-track skip~\cite{yang2012exploiting}, to identify potential noisy interactions~\cite{liu2010understanding,kim2014modeling,wen2019leveraging,jiang2020aspect} or to include such signals directly into the recommendation model~\cite{wen2019leveraging}.Such approaches, however, require additional data collection~\cite{wang2021denoising}. In addition, prior work on denoising has focused primarily on \emph{in-training} denoising, inspired from robust learning~\cite{han2018co,jiang2018mentornet} which shows that noisy signals are harder to model. To counter this, \emph{in-training} denoising re-weighs specific interactions while training the recommender model~\cite{gantner2012personalized,wang2021denoising,kaplan2021unbiased,wang2022learning}, based on item popularity, training sample loss value or prediction score. This results in custom objective functions, thereby changing the training procedure of RSs~\cite{wang2021denoising}, or training auxiliary models for noisy interactions along with the RSs model itself.~\cite{wang2022learning,lin2023autodenoise,lin2023self}. All of this is costly. 

Recently, Large Language Models (LLMs) have been used for denoising, fueled by studies showing that LLMs appear in some degree to `understand' user preferences~\cite{hou2024large,deldjoo2024review,zhao2024recommender}. This makes LLMs suitable for detecting noise in user profiles without needing additional user/item information. Prior work in this direction~\cite{wang2025unleashing,song2024large,wang2025llm4dsr} uses LLMs for denoising, but solely for in-training denoising. Wang et. al~\cite{wang2025llm4dsr} 
 first replace items in user profiles with random items, and then fine-tune an LLM to identify the random interactions in the user profiles. In addition, they use the LLM's generative capabilities to replace the noisy interactions with new ones (artificially generated). However, randomly inserting items in user profiles can lead to user profiles that no longer represent true user preferences. Fine-tuning LLMs is also much costlier than simply prompting LLMs. Lastly, using LLMs to generate interactions for users can result in out-of-catalogue items being added to user profiles. Song et. al~\cite{song2024large} propose LLMHD, where the LLM gives a score for the noise of an interaction, based on the user profile and the textual representation of the item. Then, this LLM noise score is combined with the loss value of the interactions during the training of the RSs model, thereby performing in-training denoising. 
 

In this work, we present a method for \emph{post-training} denoising of user profiles with LLMs for Collaborative Filtering (CF). Given an original (and potentially noisy) user profile constructed from implicit interactions and a fully trained recommendation model, our post-training denoising consists of changing only the input (user profile) to the trained model in order to improve the effectiveness of the model inference. In contrast to in-training denoising, this setup does not require any modification to the RSs model architecture, objective function or training procedure. Instead, we provide LLMs with (i) a user profile (past interactions), (ii) a candidate item, and (iii) its rank as given by the trained recommendation model. Then, we prompt the LLM to remove items from the user profile with an instruction to improve the rank of the candidate item. Unlike prior work on implicit feedback denoising, no supplementary interaction data are utilized, and no fine-tuning of the LLMs is performed. Our methodology relies solely on the world knowledge embedded in the pretrained LLM parameters. 
Experiments with MultiVAE~\cite{liang2018variational} -- a state-of-the-art CF model -- and 4 LLMs, both open and closed source, in 3 well-known publicly available recommendation datasets, show that our denoising approach improves recommendation effectiveness across all users, and up to 13\% for those users that are denoised. The improvements are consistent across datasets and model architectures. In summary, we contribute the first LLM-based post-training denoising approach for CF with consistent recommendation effectiveness gains.

\section{LLM-based Post-Training Denoising Methodology} \label{sec:methodology}

\begin{figure}[htbp]
    \centering
    \includegraphics[width=0.95\linewidth]{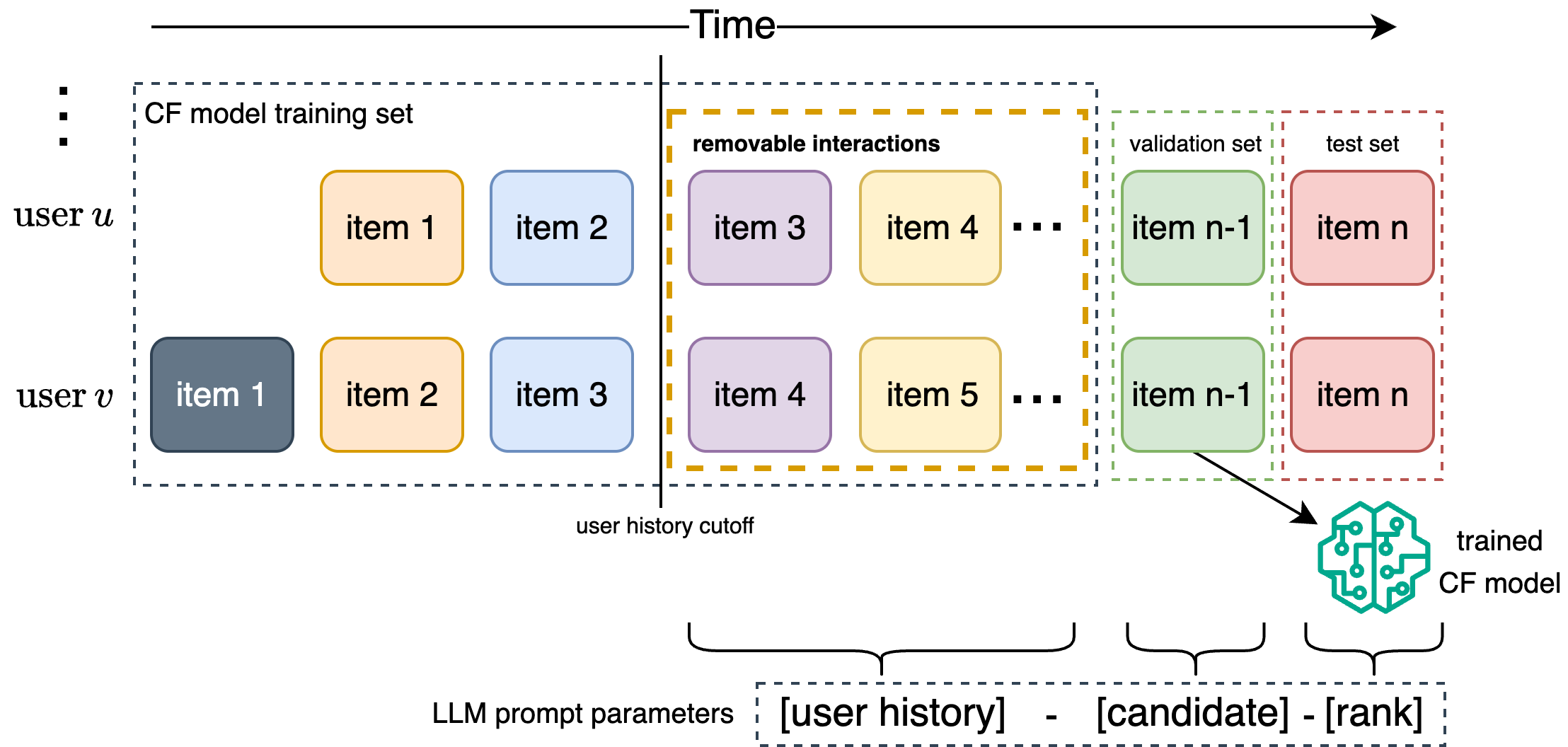}
    \caption{Post-training denoising with LLMs. The CF model is trained with the entire historical interactions of the users from the training set. The most recent interaction of each user is used as a test set. The second-to-most-recent interaction is used as validation set. The prompt of the LLM is constructed by the remaining most recent interactions of the user with the validation set item acting as a candidate item. The rank of the candidate item is retrieved by the fully trained CF model.}
    \label{fig:denoising_LLM}
\end{figure}

Given an RS dataset $\mathcal{D}$ of a set of users $U$ and a set of items $I$  -- with $\mathcal{D}_{ui} = 1$ if the user $u$ interacts with item $i$ and 0 otherwise --, and a CF model $\mathcal{M}$ fully trained on $\mathcal{D}$, post-training denoising aims to remove noisy interactions from a user profile $I_u$ in order to improve the effectiveness of $\mathcal{M}$, \emph{without retraining} $\mathcal{M}$. In order to identify the noisy interactions to remove, we convert the denoising task into a language modeling one~\cite{zhao2024recommender,hou2024large,deldjoo2024review} as follows. We provide to an LLM (i) the user profile, (ii) a candidate item from the validation set of the dataset, and (iii) the candidate's rank by the trained CF model (\cref{fig:denoising_LLM}), and prompt the LLM to remove interaction(s) from the user profile such that the candidate item's ranking by the CF model should improve. 
In a zero-shot scenario, the LLM is prompted with only the above three components (i)-(iii). In a few-shot scenario, in addition to components (i) - (iii), we use In-Context-Learning (ICL)~\cite{wang2024towards,bao2025customizing,zhao2024recommender,hou2024large,lin2025can} and provide to the LLM one of following examples: a) top-10 recommendations from the CF model, or b) items detected as noisy and their effect on another candidate item\footnote{Note that we \textbf{do not} use any additional information regarding the interaction, users or items, for these examples. The examples are built directly from the implicit feedback.} (We show how we build examples of (b) in~\cref{sec:experiments}).
More formally:

\begin{equation}
    \begin{aligned}[c]
        r_j &= \mathcal{M}(I_u, j) \\
        k &= LLM(I_u, j, r_j, e) \quad \text{s.t} \quad k \in I_u \\
        r^\prime_j &= \mathcal{M}(I_u\setminus k, j)
    \end{aligned}
\end{equation}
where $j$ is the candidate item, $r_j$ is its rank as predicted by $\mathcal{M}$ using the original user profile $I_u$, $e$ are the (optional) examples, $k$ the item to be removed from $I_u$ and $r^\prime_j$ is the candidate's updated rank after denoising the user profile (but without retraining $\mathcal{M}$). The precise prompt\footnote{In this prompt we ask the LLM to remove only 1 item. We experiment also with removing 2 items from user profiles.} we use for the LLM is shown below:


\begin{mdframed}
\scriptsize
You will help in cleaning the user historical interactions in the context of a recommender system in the movie domain. Given the user history, a candidate item and its rank (lower rank is better) by MultiVAE in the format:

\vspace{1em}

\noindent [user history] - [candidate] - [rank]

\vspace{1em}

\noindent remove only 1 item from the user history that would make MultiVAE rank the respective candidate lowest/best. List only the item to be removed from the user history in the format [item1] and nothing else. The removed item must be present in the user history.

\vspace{1em}
\noindent[Die Hard, Back to the Future, Home Alone, Toy Story, ...] - [Lion King] - [13]

\vspace{1em}

\noindent Removal:
\end{mdframed}

In the prompt, we use a candidate item from the validation set (not the test set) because this grounds the denoising process for the LLM to an item that the user has interacted with, which in turn helps the denoising be more generalizable (see~\cref{sec:results}). Moreover, the candidate's rank by the CF model indicates to the LLM how much space there is for improvement for correctly ranking the candidate item at the top of the recommendation list.

After removing the items detected by the LLM from the user profile, we run the CF model on the \emph{clean} user profile and measure the change in rank of the validation set candidate items \textbf{without} re-training the CF model. If the removal of the noisy items does not improve the rank of the candidate item, we do not remove these identified noisy items from the user profile (this happens ca. 50\% of the time, see \cref{sec:results} for details). In this case, we use the original user profile. Otherwise, we use the denoised user profile. Finally 
we run the CF model on these resulting user profiles and report the effectiveness on the test set. 

\section{Experimental Evaluation} \label{sec:experiments}
Our experiments aim to answer these three research questions:
\textbf{RQ1}: How does our LLM-based post-training denoising compare in recommendation effectiveness to no-denoising and to other post-training denoising baselines?
\textbf{RQ2}: How do different prompt formulations affect the performance of post-training denoising with LLMs?
\textbf{RQ3}: What types of items and users are more likely to benefit from our denoising method? 
Next we present our experimental setup.

\subsubsection{Evaluation Measures} \label{sec:measures}
We measure recommendation effectiveness with Normalized Discounted Cumulative Gain (NDCG)~\cite{jarvelin2002cumulated}, Hit Rate and MRR, at cutoffs 10, 20 and 100. 
We report the statistical significance between the results of each denoising method and no-denoising (original user profiles) using a paired t-test. For different LLMs and different prompt variations, we also report the percentage of \emph{formatting} errors (the LLM fails to follow the correct formatting for the output) and \emph{hallucinations} errors (the interaction(s) identified as noisy by the LLM are not in the user profile) across users.

\subsubsection{Datasets} \label{sec:datasets}
\Cref{tab:dataset_stats} displays the statistics of our datasets. 
We use MovieLens 1M~\cite{harper2015movielens}, Yelp\footnote{\url{https://www.yelp.com/dataset}} and Amazon CDs \& Vinyl~\cite{ni2019justifying}. All ratings are binarized by setting ratings greater than 0 to 1. Since we use the leave-one-out evaluation strategy, we order the interactions of each user based on the interaction timestamp\footnote{Our method does not model the interaction ordering in the user profiles, so it can be applied to non ordered interactions too.}. Then, we reserve the most recent interaction in the test set and the second-to-most-recent interaction as the validation set (used to build the LLM prompts).
Finally, we train the CF model with the remaining implicit feedback. For Yelp and Amazon CDs \& Vinyl, we filter users and items to 20-core and 10-core, respectively. 
For computational efficiency, we run the experiments on a sample of 10000 users for Yelp and Amazon CDs \& Vinly, stratifying by the number of interactions. Since LLMs can be negatively affected by a high number of user interactions in the prompt~\cite{hou2024large}, we limit the number of items that represent the user profile in the prompt to the minimum number of interactions of a user in the training set.

\subsubsection{Collaborative Filtering Model} \label{sec:cf_model}
We use 
MultiVAE~\cite{liang2018variational}, a state-of-the-art CF model based on variational autoencoders (VAE)~\cite{kingma2013auto}. We use RecBole\footnote{\url{https://recbole.io/}} to train MultiVAE and use the default hyperparameters as given in the paper~\cite{liang2018variational}. MultiVAE takes as input the binarized interactions of a user, learns a user representation in its bottleneck layer, and finally reconstructs the user interactions in its output. Items are ranked based on the output scores predicted by MultiVAE and then they are recommended to the user. Given a fully trained MultiVAE and a denoised user profile with removed interactions, we flip the input corresponding to the removed interaction(s) and run a forward pass with MultiVAE to get updated scores over the entire set of items. Our code will be made publicly available.

\begin{table}[htbp]
    \small
    \centering
    \caption{Dataset statistics after pre-processing.}
    \begin{tabular}{|l|r|r|r|}
         \hline
         & MovieLens 1M & Yelp & Amazon CDs \& Vinyl \\
         \hline
         Interactions & \num{1000209} & \num{476769} & \num{242932} \\
         Users & \num{6040} & \num{10000} & \num{10000} \\
         Items & \num{3883} & \num{20073} & \num{15180} \\
         Min. interactions per user & 20 & 20 & 10 \\
         Min. interactions per item & 1 & 20 & 10 \\
         Sparsity & 95.74\% & 99.76\% & 99.84\%\\
         \hline
    \end{tabular}
    \label{tab:dataset_stats}
\end{table}

\subsubsection{Large Language Models} \label{sec:llms}
We use the following 2 open source and 2 closed source LLMs: 
\textit{\textbf{1. Qwen 3:}~\cite{yang2025qwen3}} this is an open source model released by Alibaba in May 2025, integrating both thinking and non-thinking modes in a single model. 
We use the 8B version of Qwen 3 in non-thinking mode\footnote{\url{https://huggingface.co/Qwen/Qwen3-8B}}.
\textit{{\textbf{2. Mistral NeMo:}}}\footnote{\url{https://mistral.ai/news/mistral-nemo}} this is a 12B parameter open source model released by Mistral and NVIDIA in July 2024. We use the in\-struc\-tion-tuned version from HuggingFace\footnote{\url{https://huggingface.co/mistralai/Mistral-Nemo-Instruct-2407}}.
\textit{\textbf{3. GPT-4.1-mini}:}\footnote{\url{https://openai.com/index/gpt-4-1/}} this is a closed source model released by OpenAI in April 2025, with improved instruction following capabilities. 
The number of parameters has not been made available by OpenAI.
\textit{\textbf{4. DeepSeek-V3.1:}}\footnote{\url{https://api-docs.deepseek.com/news/news250821}}~\cite{liu2024deepseek} this is a 670B parameter open source model released by DeepSeek in August 2025, combining thinking and non-thinking modes in a single model. Similar to Qwen 3, we use the non-thinking version.

 We use the HuggingFace transformers library\footnote{\url{https://huggingface.co/docs/transformers/en/index}} as the inference engine for all open source LLMs\footnote{Denoising all \num{10000} users on Yelp takes 2 hours with an NVIDIA H100 GPU.} and OpenAI Python SDK\footnote{\url{https://github.com/openai/openai-python}} to run API-based inference for GPT-4.1-mini and DeepSeek-V3.1. For each LLM and each prompt, we perform 3 denoising runs -- to account for stochastic decoding on the LLM\footnote{We use the default decoding configuration (temperature, top-p, top-k) of the LLMs.} --, and report the mean of the runs. We used regular expressions and some light manual editing (details in the code) to clean the output of the LLMs. We also experimented with Llama 3.1 8B, however the output required extensive cleaning.

\subsubsection{Prompt variations}
We use the following prompt variations:


\noindent\textit{\textbf{Zero-shot:}} we prompt the LLM with the user profile, candidate item and its rank as given by the CF model.

\noindent\textit{\textbf{Few-shot:}} on a second validation set (third-to-most-recent interaction), for each user we remove interactions one at a time from their profile and observe the rank change of the candidates in this second validation set. We provide to the LLM the best and worst removals as denoising examples.

\noindent\textit{\textbf{Zero-shot with recommendations:}} we provide as examples the top-10 recommendations from the CF model to the LLM, which can be used to infer how to perform the denoising. As an example, consider a candidate item whose rank in the list is 8. In order to push the candidate item higher up in the ranked list, the LLM might consider removing items from user's profile that are similar to recommendations in positions 1-7 but different from the candidate item, aligning the CF model more with the given candidate.
\subsubsection{Baselines} \label{sec:baselines}
Our denoising method is the only post-training denoising method of user profiles in the literature. 
All prior work on denoising is in-training (see~\cref{sec:intro}) and is therefore not comparable to our method, because it would require retraining the CF model. We therefore use the following baselines:

\noindent\textit{\textbf{Random:}} we randomly remove an item from the most recent training interactions of each user. This is similar to masking individual items in test-time augmentation~\cite{dang2025data}.

\noindent\textit{\textbf{Top-popular:}} prior work on in-training denoising uses item popularity as a proxy for re-weighing interactions during the training of the RS~\cite{gantner2012personalized}. Based on this, we include a top-popular baseline where for each user, we remove their most popular item from the most recent training interactions.

\noindent\textit{\textbf{Semantic:}} for each user, we remove the item with the lowest cosine similarity to the user profile. We compute the cosine similarity between the item embeddings and user embeddings (mean over all user's interactions) derived from the CF model.

\noindent\textit{\textbf{UpperBoundOnVal:}} for each user, we remove every item from their most recent interactions, one at a time and observe the change in rank of the candidate item (from the validation set) as given by the CF model. The removal with the highest improvement of the candidate rank is removed from the user profile. This baseline is an upper bound in the validation set.

\noindent Since user implicit feedback might contain more than one noisy interaction, for all baselines and LLM-based denoising approaches we experiment with removing also combinations of 2 items from the most recent interactions of each user\footnote{We refrain from removing more items from the users profiles due to LLMs' proneness to hallucinations~\cite{wang2025unleashing}.}.

\section{Results} \label{sec:results}

\begin{table}[htbp]
    \centering
    \caption{Relative change (in \%) over original (not denoised) user profiles \textbf{across all users}. Best scores are in bold, and second-best are underlined. \colorbox{Cyan!50}{Blue} marks scores better than at least one \emph{random} method.  \colorbox{Orange!50}{Orange} marks scores better than at least one \emph{upperBoundOnVal} method. Darker shades mark better results. Columns For. and Hal. show the portion of LLM formatting errors and hallucinations, respectively. Statistical significance is given by the paired t-test. 
    }
    \resizebox{\textwidth}{!}{
    \begin{tabular}{|c|c|c|c|c|c|c|c|c|c|c|c|c|c|}
        \hline
        \multirow{2}{*}{} & \multirow{2}{*}{LLM} & \multirow{2}{*}{Method} & \multicolumn{3}{c|}{NDCG} & \multicolumn{3}{c|}{HR} & \multicolumn{3}{c|}{MRR} & \multicolumn{2}{c|}{Errors} \\
        \cline{4-14}
        & & & @10 & @20 & @100 & @10 & @20 & @100 & @10 & @20 & @100 & For. & Hal. \\
        
        \hline

        \multirow{16}{*}{\rotatebox{90}{MovieLens 1M}} & \multirow{8}{*}{-} & random-1 & 0.1 & -0.2 & 0.3 & 0.1 & -0.4 & 0.3 & 0.0 & -0.0 & 0.1 & \multirow{8}{*}{-} & \multirow{8}{*}{-} \\
        & & random-2 & 0.1 & -0.3 & 0.2 & 0.3 & -0.5 & 0.2 & -0.1 & -0.2 & 0.0 & & \\
        & & top-pop-1 & 1.4 & 0.9 & 0.3 & 2.6$^{*}$ & 1.4 & 0.2 & 0.4 & 0.2 & 0.2 & & \\
        & & top-pop-2 & 0.6 & 1.7$^{*}$ & 0.8$^{\dagger}$ & 0.9 & 2.6$^{\dagger}$ & 0.8$^{*}$ & 0.2 & 0.8 & 0.6 & & \\
        & & semantic-1 & -0.1 & -0.5 & 0.1 & 0.0 & -0.7 & 0.2 & -0.3 & -0.4 & -0.2 & & \\
        & & semantic-2 & 0.3 & -0.0 & 0.2 & 0.6 & 0.0 & 0.3 & -0.0 & -0.1 & -0.0 & & \\
        & & upperBoundOnVal-1 & 1.5 & 2.0$^{\dagger}$ & 1.0$^{\ddagger}$ & 1.7 & 2.4$^{*}$ & 0.6$^{*}$ & 1.3 & 1.6 & 1.3 & & \\
        & & upperBoundOnVal-2 & \textbf{3.0}$^{\dagger}$ & \textbf{3.1}$^{\ddagger}$ & \textbf{2.1}$^{\ddagger}$ & \textbf{3.7}$^{*}$ & \textbf{3.6}$^{\dagger}$ & \textbf{1.8}$^{\ddagger}$ & \textbf{2.4} & \textbf{2.5}$^{*}$ & \textbf{2.4}$^{*}$ & & \\

        \hhline{|~|*{13}{-}}
        
        & \multirow{2}{*}{Qwen} & 0-shot-recs-2 & \cellcolor{Cyan!40}1.4 & \cellcolor{Cyan!40}1.5 & \cellcolor{Cyan!40}0.9 & \cellcolor{Orange!27}2.0 & \cellcolor{Cyan!40}1.9$^{\dagger}$ & \cellcolor{Orange!40}\underline{0.9} & \cellcolor{Cyan!40}0.9 & \cellcolor{Cyan!40}1.0 & \cellcolor{Cyan!40}0.8 & 8.3\% & 9.1\% \\
        & & few-shot-2 & \cellcolor{Orange!40}\underline{2.3} & \cellcolor{Orange!40}\underline{2.8}$^{*}$ & \cellcolor{Orange!40}\underline{1.3}$^{\dagger}$ & \cellcolor{Orange!40}2.3 & \cellcolor{Orange!31}\underline{2.9}$^{\dagger}$ & \cellcolor{Orange!40}0.8$^{\dagger}$ & \cellcolor{Orange!40}\underline{2.3} & \cellcolor{Orange!40}\underline{2.5} & \cellcolor{Orange!40}\underline{2.0} & 1.1\% & 0.7\% \\
        
        \hhline{|~|*{13}{-}}
        
        & \multirow{2}{*}{Mistral} & 0-shot-recs-2 & \cellcolor{Cyan!40}0.5 & \cellcolor{Cyan!40}0.2 & \cellcolor{Cyan!40}0.6 & \cellcolor{Cyan!40}0.6 & \cellcolor{Cyan!40}0.0$^{\dagger}$ & \cellcolor{Orange!35}0.8 & \cellcolor{Cyan!40}0.3 & \cellcolor{Cyan!40}0.2 & \cellcolor{Cyan!40}0.4 & 1.2\% & 20.1\% \\
        & & few-shot-2 & \cellcolor{Orange!40}2.3 & \cellcolor{Cyan!40}1.7$^{*}$ & \cellcolor{Orange!15}1.0$^{\dagger}$ & \cellcolor{Orange!40}\underline{3.2} & \cellcolor{Cyan!40}1.9$^{\dagger}$ & \cellcolor{Orange!40}\underline{0.9}$^{\dagger}$ & \cellcolor{Orange!40}1.7 & \cellcolor{Cyan!40}1.4 & \cellcolor{Cyan!40}1.2 & 3.2\% & 3.2\% \\
        
        \hhline{|~|*{13}{-}}
        
        & \multirow{2}{*}{GPT} & 0-shot-recs-2 & \cellcolor{Cyan!40}0.6 & \cellcolor{Cyan!40}1.2 & \cellcolor{Cyan!40}0.5 & \cellcolor{Cyan!40}0.5 & \cellcolor{Cyan!40}1.5$^{\dagger}$ & \cellcolor{Cyan!40}0.3 & \cellcolor{Cyan!40}0.7 & \cellcolor{Cyan!40}0.9 & \cellcolor{Cyan!40}0.7 & 0.2\% & 1.6\% \\
        & & few-shot-2 & \cellcolor{Cyan!40}0.4 & \cellcolor{Cyan!40}0.9$^{*}$ & \cellcolor{Cyan!40}0.4$^{\dagger}$ & \cellcolor{Cyan!40}0.6 & \cellcolor{Cyan!40}1.3$^{\dagger}$ & \cellcolor{Cyan!40}0.4$^{\dagger}$ & \cellcolor{Cyan!40}0.3 & \cellcolor{Cyan!40}0.6 & \cellcolor{Cyan!40}0.4 & 0.1\% & 1.2\% \\
        
        \hhline{|~|*{13}{-}}
        
        & \multirow{2}{*}{DeepSeek} & 0-shot-recs-2 & -0.1 & \cellcolor{Cyan!40}0.9 & \cellcolor{Cyan!40}0.3 & \cellcolor{Cyan!40}0.5 & \cellcolor{Cyan!40}2.0$^{\dagger}$ & \cellcolor{Cyan!40}0.4 & -0.7 & -0.2 & -0.3 & 0.0\% & 13.4\% \\
        & & few-shot-2 & \cellcolor{Cyan!40}0.2 & \cellcolor{Cyan!40}1.5$^{*}$ & \cellcolor{Cyan!40}0.7$^{\dagger}$ & \cellcolor{Cyan!40}0.7 & \cellcolor{Orange!20}2.6$^{\dagger}$ & \cellcolor{Orange!40}0.8$^{\dagger}$ & -0.2 & \cellcolor{Cyan!40}0.4 & \cellcolor{Cyan!40}0.3 & 0.0\% & 1.0\% \\
        
        \hline
        \hline

        \multirow{16}{*}{\rotatebox{90}{Yelp}} & \multirow{8}{*}{-} & random-1 & 0.1 & 0.4 & 0.0 & 0.5 & 0.8 & 0.0 & -0.2 & 0.0 & -0.1 & \multirow{8}{*}{-} & \multirow{8}{*}{-} \\
        & & random-2 & 0.1 & 0.2 & 0.2 & -0.2 & -0.1 & 0.1 & 0.4 & 0.4 & 0.5 & & \\
        & & top-pop-1 & -0.1 & 0.7 & 0.7$^{\dagger}$ & 0.0 & 1.0 & 0.7$^{*}$ & -0.1 & 0.3 & 0.4 & & \\
        & & top-pop-2 & 1.4 & \underline{2.0}$^{\dagger}$ & 1.0$^{\dagger}$ & 1.5 & \textbf{2.5}$^{*}$ & 0.8$^{*}$ & 1.2 & 1.5$^{*}$ & 1.3$^{*}$ & & \\
        & & semantic-1 & 1.5$^{*}$ & -0.0 & 0.3 & 2.2$^{*}$ & -0.6 & 0.2 & 1.0$^{*}$ & 0.5 & 0.6$^{\dagger}$ & & \\
        & & semantic-2 & 1.7$^{*}$ & 0.1 & 0.6$^{*}$ & 2.2$^{*}$ & -0.6 & 0.4 & 1.4 & 0.7 & 0.9 & & \\
        & & upperBoundOnVal-1 & \underline{2.6}$^{*}$ & 1.6 & 1.1$^{\ddagger}$ & \underline{2.9}$^{*}$ & 1.2 & 0.9$^{*}$ & \underline{2.4}$^{*}$ & 2.0$^{*}$ & \underline{1.8}$^{\dagger}$ & & \\
        & & upperBoundOnVal-2 & \textbf{5.0}$^{\dagger}$ & \textbf{3.4}$^{\dagger}$ & \textbf{2.3}$^{\ddagger}$ & \textbf{4.7}$^{*}$ & \underline{2.3} & \textbf{1.6}$^{\dagger}$ & \textbf{5.5}$^{\dagger}$ & \textbf{4.7}$^{\dagger}$ & \textbf{4.1}$^{\dagger}$ & & \\
        
        \hhline{|~|*{13}{-}}
        
        & \multirow{2}{*}{Qwen} & 0-shot-recs-2 & \cellcolor{Cyan!40}0.5 & \cellcolor{Cyan!40}1.4 & \cellcolor{Cyan!40}0.2 & \cellcolor{Cyan!40}0.4 & \cellcolor{Orange!40}1.9 & -0.1 & \cellcolor{Cyan!40}0.7 & \cellcolor{Cyan!40}1.1 & \cellcolor{Cyan!40}0.7 & 11.8\% & 12.2\% \\
        & & few-shot-2 & \cellcolor{Cyan!40}1.9$^{*}$ & \cellcolor{Orange!27}1.8 & \cellcolor{Orange!19}\underline{1.2}$^{*}$ & \cellcolor{Cyan!40}1.5 & \cellcolor{Orange!40}1.7 & \cellcolor{Orange!24}\underline{1.0} & \cellcolor{Cyan!40}2.1 & \cellcolor{Orange!12}\underline{2.0} & \cellcolor{Cyan!40}1.7 & 0.2\% & 1.6\% \\
        
        \hhline{|~|*{13}{-}}
        
        & \multirow{2}{*}{Mistral} & 0-shot-recs-2 & \cellcolor{Cyan!40}0.4 & \cellcolor{Cyan!40}0.6 & \cellcolor{Cyan!40}0.5 & \cellcolor{Cyan!40}0.4 & \cellcolor{Cyan!40}0.6 & \cellcolor{Cyan!40}0.5 & \cellcolor{Cyan!40}0.4 & \cellcolor{Cyan!40}0.5 & \cellcolor{Cyan!40}0.4 & 0.6\% & 37.2\% \\
        & & few-shot-2 & \cellcolor{Cyan!40}0.9$^{*}$ & 0.1 & \cellcolor{Cyan!40}0.5$^{*}$ & \cellcolor{Cyan!40}1.1 & -0.4 & \cellcolor{Cyan!40}0.4 & \cellcolor{Cyan!40}0.7 & \cellcolor{Cyan!40}0.5 & \cellcolor{Cyan!40}0.6 & 4.0\% & 8.1\% \\
        
        \hhline{|~|*{13}{-}}
        
        & \multirow{2}{*}{GPT} & 0-shot-recs-2 & \cellcolor{Cyan!40}0.5 & \cellcolor{Cyan!40}0.3 & \cellcolor{Cyan!40}0.2 & \cellcolor{Cyan!40}0.6 & \cellcolor{Cyan!40}0.3 & \cellcolor{Cyan!40}0.1 & \cellcolor{Cyan!40}0.4 & \cellcolor{Cyan!40}0.4 & \cellcolor{Cyan!40}0.4 & 0.1\% & 5.0\% \\
        & & few-shot-2 & \cellcolor{Cyan!40}0.3$^{*}$ & -0.1 & \cellcolor{Cyan!34}0.0$^{*}$ & \cellcolor{Cyan!40}0.5 & -0.2 & -0.0 & \cellcolor{Cyan!40}0.2 & 0.0 & \cellcolor{Cyan!40}0.1 & 0.1\% & 4.1\% \\
        
        \hhline{|~|*{13}{-}}
        
        & \multirow{2}{*}{DeepSeek} & 0-shot-recs-2 & \cellcolor{Cyan!40}0.4 & \cellcolor{Cyan!40}0.4 & \cellcolor{Cyan!40}0.4 & \cellcolor{Cyan!40}0.4 & \cellcolor{Cyan!40}0.2 & \cellcolor{Cyan!40}0.3 & \cellcolor{Cyan!40}0.6 & \cellcolor{Cyan!40}0.6 & \cellcolor{Cyan!40}0.6 & 0.0\% & 19.4\% \\
        & & few-shot-2 & \cellcolor{Cyan!40}1.8$^{*}$ & \cellcolor{Cyan!40}1.2 & \cellcolor{Cyan!40}0.8$^{*}$ & \cellcolor{Cyan!40}1.5 & \cellcolor{Cyan!40}0.7 & \cellcolor{Cyan!40}0.4 & \cellcolor{Cyan!40}2.0 & \cellcolor{Cyan!40}1.7 & \cellcolor{Cyan!40}1.5 & 0.0\% & 4.8\% \\
        
        \hline
        \hline

        \multirow{16}{*}{\rotatebox{90}{Amazon CDs \& Vinyl}} & \multirow{8}{*}{-} & random-1 & 1.1 & 1.3 & 1.2$^{\dagger}$ & 0.1 & 0.7 & 1.0$^{*}$ & 1.9 & 1.9 & 1.8 & \multirow{8}{*}{-} & \multirow{8}{*}{-} \\
        & & random-2 & 2.5$^{*}$ & 1.5 & 0.8 & 2.5 & 0.9 & 0.1 & 2.6 & 2.1 & 1.9 & & \\
        & & top-pop-1 & -0.3 & 0.5 & 0.3 & -0.4 & 1.0 & 0.5 & 0.0 & 0.3 & 0.3 & & \\
        & & top-pop-2 & -1.7 & 0.5 & -0.2 & -1.9 & 2.1 & 0.3 & -1.6 & -0.7 & -0.9 & & \\
        & & semantic-1 & 1.2 & 1.1 & 0.6 & 0.6 & 0.7 & 0.2 & 1.7 & 1.6 & 1.5 & & \\
        & & semantic-2 & 0.4 & 1.0 & 0.3 & 0.8 & 1.6 & 0.2 & 0.1 & 0.5 & 0.4 & & \\
        & & upperBoundOnVal-1 & \underline{4.7}$^{*}$ & 2.6 & \underline{2.5}$^{\dagger}$ & \underline{7.6}$^{\dagger}$ & 3.0 & \underline{2.4}$^{\dagger}$ & 2.6 & 1.9 & 2.1 & & \\
        & & upperBoundOnVal-2 & \textbf{13.4}$^{\ddagger}$ & \textbf{11.5}$^{\ddagger}$ & \textbf{7.3}$^{\ddagger}$ & \textbf{12.3}$^{\ddagger}$ & \textbf{9.4}$^{\ddagger}$ & \textbf{4.1}$^{\ddagger}$ & \textbf{14.2}$^{\ddagger}$ & \textbf{13.4}$^{\ddagger}$ & \textbf{12.0}$^{\ddagger}$ & & \\
        
        \hhline{|~|*{13}{-}}
        
        & \multirow{2}{*}{Qwen} & 0-shot-recs-2 & \cellcolor{Cyan!40}2.2$^{\dagger}$ & \cellcolor{Cyan!40}1.6$^{\dagger}$ & 0.7$^{*}$ & \cellcolor{Cyan!40}2.9$^{\dagger}$ & \cellcolor{Cyan!40}1.7$^{*}$ & -0.2 & 1.6$^{*}$ & 1.5$^{*}$ & 1.4$^{*}$ & 7.2\% & 11.6\% \\
        & & few-shot-2 & \cellcolor{Cyan!40}3.0$^{\dagger}$ & \cellcolor{Cyan!40}2.4$^{\ddagger}$ & \cellcolor{Cyan!40}2.1$^{\ddagger}$ & \cellcolor{Cyan!40}2.7$^{*}$ & \cellcolor{Cyan!40}1.6$^{\dagger}$ & \cellcolor{Cyan!40}1.4$^{*}$ & \cellcolor{Orange!38}3.4$^{*}$ & \cellcolor{Orange!40}3.2$^{\dagger}$ & \cellcolor{Orange!40}3.2$^{\dagger}$ & 5.0\% & 2.7\% \\
        
        \hhline{|~|*{13}{-}}
        
        & \multirow{2}{*}{Mistral} & 0-shot-recs-2 & \cellcolor{Cyan!40}4.2$^{\dagger}$ & \cellcolor{Orange!26}3.0$^{\dagger}$ & \cellcolor{Cyan!40}2.1$^{*}$ & \cellcolor{Cyan!40}4.7$^{\dagger}$ & \cellcolor{Cyan!40}2.6$^{*}$ & \cellcolor{Cyan!40}1.6 & \cellcolor{Orange!40}4.0$^{*}$ & \cellcolor{Orange!40}3.5$^{*}$ & \cellcolor{Orange!40}3.2$^{*}$ & 4.1\% & 24.2\% \\
        & & few-shot-2 & \cellcolor{Cyan!40}2.4$^{\dagger}$ & \cellcolor{Orange!40}3.6$^{\ddagger}$ & \cellcolor{Cyan!40}1.7$^{\ddagger}$ & \cellcolor{Cyan!40}3.0$^{*}$ & \cellcolor{Orange!40}\underline{4.9}$^{\dagger}$ & \cellcolor{Cyan!40}1.3$^{*}$ & \cellcolor{Cyan!14}1.9$^{*}$ & \cellcolor{Orange!37}2.5$^{\dagger}$ & \cellcolor{Cyan!17}2.0$^{\dagger}$ & 9.6\% & 10.2\% \\
        
        \hhline{|~|*{13}{-}}
        
        & \multirow{2}{*}{GPT} & 0-shot-recs-2 & \cellcolor{Cyan!40}3.3$^{\dagger}$ & \cellcolor{Orange!39}3.4$^{\dagger}$ & \cellcolor{Cyan!40}1.5$^{*}$ & \cellcolor{Cyan!40}2.6$^{\dagger}$ & \cellcolor{Orange!14}3.1$^{*}$ & \cellcolor{Cyan!40}0.4 & \cellcolor{Orange!40}3.8$^{*}$ & \cellcolor{Orange!40}3.8$^{*}$ & \cellcolor{Orange!40}3.2$^{*}$ & 0.6\% & 2.7\% \\
        & & few-shot-2 & \cellcolor{Cyan!31}1.3$^{\dagger}$ & \cellcolor{Cyan!34}1.6$^{\ddagger}$ & 0.8$^{\ddagger}$ & \cellcolor{Cyan!40}0.8$^{*}$ & \cellcolor{Cyan!40}1.4$^{\dagger}$ & \cellcolor{Cyan!40}0.2$^{*}$ & 1.7$^{*}$ & 1.8$^{\dagger}$ & 1.6$^{\dagger}$ & 0.1\% & 2.5\% \\
        
        \hhline{|~|*{13}{-}}
        
        & \multirow{2}{*}{DeepSeek} & 0-shot-recs-2 & \cellcolor{Cyan!40}4.4$^{\dagger}$ & \cellcolor{Orange!31}3.2$^{\dagger}$ & \cellcolor{Cyan!40}1.5$^{*}$ & \cellcolor{Cyan!40}4.3$^{\dagger}$ & \cellcolor{Cyan!40}2.3$^{*}$ & 0.1 & \cellcolor{Orange!40}4.5$^{*}$ & \cellcolor{Orange!40}4.0$^{*}$ & \cellcolor{Orange!40}3.4$^{*}$ & 0.0\% & 4.6\% \\
        & & few-shot-2 & \cellcolor{Cyan!40}3.7$^{\dagger}$ & \cellcolor{Orange!40}\underline{3.6}$^{\ddagger}$ & \cellcolor{Cyan!40}2.3$^{\ddagger}$ & \cellcolor{Cyan!40}3.0$^{*}$ & \cellcolor{Orange!14}3.1$^{\dagger}$ & \cellcolor{Cyan!40}1.2$^{*}$ & \cellcolor{Orange!40}\underline{4.5}$^{*}$ & \cellcolor{Orange!40}\underline{4.4}$^{\dagger}$ & \cellcolor{Orange!40}\underline{4.0}$^{\dagger}$ & 0.0\% & 2.1\% \\
        
        \hline
        
        \multicolumn{14}{l}{$^\ddagger \, p < 0.001$, $^\dagger \, p < 0.01$, $^* \, p < 0.05$}
    \end{tabular}
    }
    \label{tab:overall_results}
\end{table}

\begin{table}[!ht]
    \centering
    \caption{Relative change (in \%) on denoised users, i.e., their profiles were cleaned. Best scores are in bold, and second-best are underlined. \colorbox{Cyan!50}{Blue} marks scores better than at least one \emph{random} method.  \colorbox{Orange!50}{Orange} marks scores better than at least one \emph{upperBoundOnVal} method. Darker shades mark better results. Columns For. and Hal. show the portion of LLM formatting errors and hallucinations, respectively. Stat. significance is given by the paired t-test. 
    }
    \resizebox{\textwidth}{!}{
    \begin{tabular}{|c|c|c|c|c|c|c|c|c|c|c|c|c|c|c|}
        \hline
        \multirow{2}{*}{} & \multirow{2}{*}{LLM} & \multirow{2}{*}{Method} & \multicolumn{3}{c|}{NDCG} & \multicolumn{3}{c|}{HR} & \multicolumn{3}{c|}{MRR} & \multirow{2}{*}{\shortstack{\% users \\ denoised}} & \multicolumn{2}{c|}{Errors} \\
        
        \cline{4-12}\cline{14-15}
        
        & & & @10 & @20 & @100 & @10 & @20 & @100 & @10 & @20 & @100 & & For. & Hal. \\
        
        \hline
        
        \multirow{10}{*}{\rotatebox{90}{MovieLens 1M}} & \multirow{2}{*}{-} & upperBoundOnVal-1 & 1.8 & 2.4$^{\dagger}$ & 1.1$^{\ddagger}$ & 2.1 & 2.8$^{*}$ & 0.7$^{*}$ & 1.5 & 1.9 & 1.6 & 92.4\% & \multirow{2}{*}{-} & \multirow{2}{*}{-} \\
        & & upperBoundOnVal-2 & 3.4$^{\dagger}$ & 3.5$^{\ddagger}$ & 2.3$^{\ddagger}$ & 4.2$^{*}$ & 4.0$^{\dagger}$ & \underline{1.9}$^{\ddagger}$ & 2.7 & 2.9$^{*}$ & 2.6$^{*}$ & 95.3\% & & \\
        
        \hhline{|~|*{14}{-}}
        
        & \multirow{2}{*}{Qwen} & 0-shot-recs-2 & \cellcolor{Orange!26}2.1 & \cellcolor{Orange!40}3.6 & \cellcolor{Orange!40}2.1$^{*}$ & \cellcolor{Orange!16}2.2 & \cellcolor{Orange!40}4.3$^{*}$ & \cellcolor{Orange!40}1.9$^{*}$ & \cellcolor{Orange!16}1.6 & \cellcolor{Orange!40}2.5 & \cellcolor{Orange!40}2.1 & 42.0\% & 8.9\% & 9.5\% \\
        & & few-shot-2 & \cellcolor{Orange!40}\textbf{6.0} & \cellcolor{Orange!40}\textbf{6.4}$^{*}$ & \cellcolor{Orange!40}\underline{2.8}$^{\dagger}$ & \cellcolor{Orange!40}\textbf{6.4} & \cellcolor{Orange!40}\textbf{6.7}$^{\dagger}$ & \cellcolor{Orange!40}1.7$^{\dagger}$ & \cellcolor{Orange!40}\textbf{5.5} & \cellcolor{Orange!40}\textbf{5.8} & \cellcolor{Orange!40}\textbf{4.5} & 50.6\% & 1.1\% & 0.6\% \\
        
        \hhline{|~|*{14}{-}}
        
        & \multirow{2}{*}{Mistral} & 0-shot-recs-2 & 1.3 & 0.4 & \cellcolor{Orange!27}1.3$^{*}$ & \cellcolor{Orange!27}2.4 & 0.5$^{*}$ & \cellcolor{Orange!40}1.6$^{*}$ & 0.2 & -0.1 & 0.3 & 37.2\% & 1.3\% & 19.9\% \\
        & & few-shot-2 & \cellcolor{Orange!40}\underline{5.2} & \cellcolor{Orange!40}\underline{4.8}$^{*}$ & \cellcolor{Orange!40}\textbf{2.8}$^{\dagger}$ & \cellcolor{Orange!40}\underline{6.1} & \cellcolor{Orange!40}\underline{5.2}$^{\dagger}$ & \cellcolor{Orange!40}\textbf{2.6}$^{\dagger}$ & \cellcolor{Orange!40}\underline{4.3} & \cellcolor{Orange!40}\underline{4.2} & \cellcolor{Orange!40}\underline{3.4} & 46.2\% & 2.8\% & 3.1\% \\
        
        \hhline{|~|*{14}{-}}
        
        & \multirow{3}{*}{GPT} & 0-shot-recs-2 & \cellcolor{Orange!20}2.0 & \cellcolor{Orange!24}2.7 & \cellcolor{Orange!16}1.2$^{*}$ & 1.2 & \cellcolor{Orange!11}2.8$^{*}$ & 0.7$^{*}$ & \cellcolor{Orange!40}2.5 & \cellcolor{Orange!40}2.7 & \cellcolor{Orange!40}2.1 & 52.3\% & 0.3\% & 1.6\% \\
        & & few-shot-2 & 1.1 & \cellcolor{Orange!28}2.8$^{*}$ & 1.1$^{\dagger}$ & 0.8 & \cellcolor{Orange!39}3.6$^{\dagger}$ & \cellcolor{Orange!30}0.8$^{\dagger}$ & 1.5 & \cellcolor{Orange!26}2.2 & \cellcolor{Orange!12}1.6 & 53.5\% & 0.1\% & 1.2\% \\
        
        \hhline{|~|*{14}{-}}
        
        & \multirow{2}{*}{DeepSeek} & 0-shot-recs-2 & 0.3 & 2.1 & 0.9$^{*}$ & 1.4 & \cellcolor{Orange!40}4.0$^{*}$ & \cellcolor{Orange!40}0.9$^{*}$ & -0.7 & 0.2 & 0.1 & 50.7\% & 0.0\% & 14.1\% \\
        & & few-shot-2 & \cellcolor{Orange!14}1.9 & \cellcolor{Orange!40}3.2$^{*}$ & \cellcolor{Orange!40}1.6$^{\dagger}$ & \cellcolor{Orange!40}2.7 & \cellcolor{Orange!40}4.6$^{\dagger}$ & \cellcolor{Orange!40}1.5$^{\dagger}$ & 1.2 & 1.8 & 1.4 & 57.5\% & 0.0\% & 1.0\% \\
        
        \hline
        \hline

        \multirow{10}{*}{\rotatebox{90}{Yelp}} & \multirow{2}{*}{-} & upperBoundOnVal-1 & 3.0$^{*}$ & 1.8 & 1.3$^{\ddagger}$ & \underline{3.4}$^{*}$ & 1.4 & 1.0$^{*}$ & 2.9$^{*}$ & 2.3$^{*}$ & 2.1$^{\dagger}$ & 91.7\% & \multirow{2}{*}{-} & \multirow{2}{*}{-} \\
        & & upperBoundOnVal-2 & \textbf{5.5}$^{\dagger}$ & 3.6$^{\dagger}$ & \underline{2.5}$^{\ddagger}$ & \textbf{5.1}$^{*}$ & 2.4 & \underline{1.7}$^{\dagger}$ & \textbf{6.1}$^{\dagger}$ & \textbf{5.1}$^{\dagger}$ & \textbf{4.4}$^{\dagger}$ & 94.8\% & & \\
        
        \hhline{|~|*{14}{-}}
        
        & \multirow{2}{*}{Qwen} & 0-shot-recs-2 & 1.0 & \cellcolor{Orange!40}3.8 & 0.4$^{*}$ & 0.9 & \cellcolor{Orange!40}\textbf{5.8} & -0.3 & 1.3 & \cellcolor{Orange!16}2.5 & 1.5 & 37.8\% & 12.3\% & 12.6\% \\
        & & few-shot-2 & \cellcolor{Orange!40}\underline{4.1}$^{*}$ & \cellcolor{Orange!40}\textbf{4.2} & \cellcolor{Orange!40}\textbf{2.6}$^{\dagger}$ & 3.3 & \cellcolor{Orange!40}\underline{4.0} & \cellcolor{Orange!40}\textbf{2.0} & \cellcolor{Orange!40}\underline{4.5} & \cellcolor{Orange!40}4.4 & \cellcolor{Orange!40}\underline{3.8} & 47.9\% & 0.2\% & 1.4\% \\
        
        \hhline{|~|*{14}{-}}
        
        & \multirow{2}{*}{Mistral} & 0-shot-recs-2 & 2.9 & \cellcolor{Orange!40}\underline{3.8} & \cellcolor{Orange!40}1.7$^{*}$ & 1.3 & \cellcolor{Orange!40}3.5 & \cellcolor{Orange!31}1.2 & \cellcolor{Orange!40}4.3 & \cellcolor{Orange!40}\underline{4.5} & \cellcolor{Orange!40}3.3 & 33.0\% & 0.5\% & 37.2\% \\
        & & few-shot-2 & \cellcolor{Orange!15}3.2$^{*}$ & 1.1 & \cellcolor{Orange!40}1.8$^{\dagger}$ & 3.1 & -0.6 & \cellcolor{Orange!40}1.5 & \cellcolor{Orange!24}3.3 & \cellcolor{Orange!21}2.6 & \cellcolor{Orange!38}2.6 & 40.3\% & 4.0\% & 7.7\% \\
        
        \hhline{|~|*{14}{-}}
        
        & \multirow{3}{*}{GPT} &  0-shot-recs-2 & 1.0 & 0.5 & 0.3$^{*}$ & 1.4 & 0.2 & 0.1 & 0.8 & 0.7 & 0.7 & 63.3\% & 0.1\% & 5.1\% \\
        & & few-shot-2 & 0.4$^{*}$ & -0.0 & 0.2$^{\dagger}$ & 0.6 & -0.2 & 0.1 & 0.3 & 0.1 & 0.3 & 64.1\% & 0.1\% & 4.0\% \\
        
        \hhline{|~|*{14}{-}}
        
        & \multirow{2}{*}{DeepSeek} & 0-shot-recs-2 & 1.0 & 1.5 & 0.8$^{*}$ & 0.8 & \cellcolor{Orange!21}1.6 & 0.5 & 1.4 & 1.5 & 1.3 & 52.5\% & 0.0\% & 19.7\% \\
        & & few-shot-2 & \cellcolor{Orange!12}3.1$^{*}$ & \cellcolor{Orange!10}1.8 & \cellcolor{Orange!22}1.5$^{\dagger}$ & 2.6 & 0.8 & 0.9 & \cellcolor{Orange!30}3.5 & \cellcolor{Orange!33}2.9 & \cellcolor{Orange!38}2.6 & 57.6\% & 0.0\% & 4.7\% \\

        \hline
        \hline

        \multirow{10}{*}{\rotatebox{90}{Amazon CDs \& Vinyl}} & \multirow{2}{*}{-} & upperBoundOnVal-1 & 5.2$^{*}$ & 2.9 & 2.7$^{\dagger}$ & 8.2$^{\dagger}$ & 3.2 & 2.5$^{\dagger}$ & 2.9 & 2.2 & 2.3 & 97.6\% & \multirow{2}{*}{-} & \multirow{2}{*}{-} \\
        & & upperBoundOnVal-2 & \textbf{14.5}$^{\ddagger}$ & \textbf{12.4}$^{\ddagger}$ & \textbf{7.8}$^{\ddagger}$ & \textbf{13.2}$^{\ddagger}$ & \underline{10.0}$^{\ddagger}$ & \textbf{4.3}$^{\ddagger}$ & \textbf{15.6}$^{\ddagger}$ & \textbf{14.7}$^{\ddagger}$ & \textbf{13.1}$^{\ddagger}$ & 97.8\% & & \\
        
        \hhline{|~|*{14}{-}}
        
        & \multirow{2}{*}{Qwen} & 0-shot-recs-2 & \cellcolor{Orange!40}7.6$^{\ddagger}$ & \cellcolor{Orange!40}5.5$^{\dagger}$ & 2.4$^{\dagger}$ & 8.1$^{\dagger}$ & \cellcolor{Orange!40}4.3$^{*}$ & -0.2 & \cellcolor{Orange!40}7.3$^{\dagger}$ & \cellcolor{Orange!40}6.4$^{\dagger}$ & \cellcolor{Orange!40}5.9$^{\dagger}$ & 45.3\% & 7.4\% & 12.2\% \\
        & & few-shot-2 & \cellcolor{Orange!13}5.3$^{\dagger}$ & \cellcolor{Orange!40}4.5$^{\ddagger}$ & \cellcolor{Orange!40}4.4$^{\ddagger}$ & 4.1$^{*}$ & 2.9$^{*}$ & \cellcolor{Orange!40}3.2 & \cellcolor{Orange!40}6.4$^{\dagger}$ & \cellcolor{Orange!40}6.1$^{\dagger}$ & \cellcolor{Orange!40}6.3$^{\dagger}$ & 48.0\% & 4.8\% & 2.7\% \\
        
        \hhline{|~|*{14}{-}}
        
        & \multirow{2}{*}{Mistral} & 0-shot-recs-2 & \cellcolor{Orange!40}\underline{11.7}$^{\ddagger}$ & \cellcolor{Orange!40}\underline{8.2}$^{\dagger}$ & \cellcolor{Orange!40}\underline{5.3}$^{\dagger}$ & \cellcolor{Orange!40}\underline{13.0}$^{\dagger}$ & \cellcolor{Orange!40}6.6$^{*}$ & \cellcolor{Orange!40}\underline{3.4} & \cellcolor{Orange!40}\underline{11.4}$^{\dagger}$ & \cellcolor{Orange!40}\underline{9.9}$^{\dagger}$ & \cellcolor{Orange!40}\underline{8.7}$^{\dagger}$ & 42.1\% & 4.3\% & 24.3\% \\
        & & few-shot-2 & \cellcolor{Orange!29}6.2$^{\dagger}$ & \cellcolor{Orange!40}8.0$^{\ddagger}$ & \cellcolor{Orange!40}3.8$^{\ddagger}$ & \cellcolor{Orange!14}8.5$^{*}$ & \cellcolor{Orange!40}\textbf{11.0}$^{*}$ & \cellcolor{Orange!32}3.0 & \cellcolor{Orange!40}4.4$^{\dagger}$ & \cellcolor{Orange!40}5.3$^{\dagger}$ & \cellcolor{Orange!40}4.2$^{\dagger}$ & 45.7\% & 9.9\% & 10.3\% \\
        
        \hhline{|~|*{14}{-}}
        
        & \multirow{2}{*}{GPT} & 0-shot-recs-2 & 4.7$^{\ddagger}$ & \cellcolor{Orange!40}4.7$^{\dagger}$ & 2.0$^{\dagger}$ & 3.7$^{\dagger}$ & \cellcolor{Orange!37}4.1$^{*}$ & 0.5 & \cellcolor{Orange!40}5.7$^{\dagger}$ & \cellcolor{Orange!40}5.5$^{\dagger}$ & \cellcolor{Orange!40}4.6$^{\dagger}$ & 72.1\% & 0.6\% & 2.8\% \\
        & & few-shot-2 & 2.3$^{\dagger}$ & 2.4$^{\ddagger}$ & 1.2$^{\ddagger}$ & 1.6$^{*}$ & 2.0$^{*}$ & 0.3 & \cellcolor{Orange!15}3.1$^{\dagger}$ & \cellcolor{Orange!40}3.1$^{\dagger}$ & \cellcolor{Orange!27}2.7$^{\dagger}$ & 70.5\% & 0.1\% & 2.5\% \\
        
        \hhline{|~|*{14}{-}}
        
        & \multirow{2}{*}{DeepSeek} & 0-shot-recs-2 & \cellcolor{Orange!40}7.9$^{\ddagger}$ & \cellcolor{Orange!40}5.5$^{\dagger}$ & 2.6$^{\dagger}$ & 7.6$^{\dagger}$ & \cellcolor{Orange!23}3.6$^{*}$ & 0.4 & \cellcolor{Orange!40}8.3$^{\dagger}$ & \cellcolor{Orange!40}7.2$^{\dagger}$ & \cellcolor{Orange!40}6.3$^{\dagger}$ & 65.8\% & 0.0\% & 4.4\% \\
        & & few-shot-2 & \cellcolor{Orange!40}7.6$^{\dagger}$ & \cellcolor{Orange!40}6.3$^{\ddagger}$ & \cellcolor{Orange!40}4.0$^{\ddagger}$ & 6.2$^{*}$ & \cellcolor{Orange!40}4.7$^{*}$ & 1.8 & \cellcolor{Orange!40}9.1$^{\dagger}$ & \cellcolor{Orange!40}8.4$^{\dagger}$ & \cellcolor{Orange!40}7.6$^{\dagger}$ & 65.3\% & 0.0\% & 2.1\% \\
        
        \hline

        \multicolumn{14}{l}{$^\ddagger \, p < 0.001$, $^\dagger \, p < 0.01$, $^* \, p < 0.05$}
    \end{tabular}
    }
    \label{tab:denoised_results}
\end{table}

\subsubsection{Effectiveness of post-training denoising (RQ1).} \label{sec:rq1}
In \cref{tab:overall_results} we report the relative change (in \%) of our LLM-based post-training denoising\footnote{Due to limited space we include only the best performing prompt variations.} and the baselines, over no-denoising (original user profiles). Note that the relative change is given \textbf{across all users}, i.e., including those for which denoising did not improve the rank of the candidate item in the validation set (their user profile was not denoised). We also show the percentage of both formatting errors and hallucinations errors for the LLM denoising methods. 
We see that our LLM-based denoising yields statistically significant gains over original user profiles; up to 2.9\% on MovieLens 1M (Qwen, few-shot-2 prompt) and up to 4.9\% on Amazon CDs \& Vinyl (Mistral, few-shot-2) in NDCG@20, with similar improvements also in HR@20. Although all LLM-based methods clearly improve over random baselines, on MovieLens 1M and Amazon CDs \& Vinyl several LLMs exceed the upperBoundOnVal-1\footnote{Note that upperBoundOnVal is an upper bound \textbf{only} on the validation set.}. On MovieLens 1M in particular, few-shot-2 prompt on Qwen comes close to upperBoundOnVal-2.  For Yelp, our approaches yielded fewer statistically significant gains over the random baselines. This could be due to the higher hallucination rates in this dataset, compared to the other two, which directly contribute to the performance of the LLMs in this task. Overall, \cref{tab:overall_results} shows that our post-training method is an effective option for denoising user profiles.

Given that there might be cases when denoising does not improve the rank of the candidate item, we also report on \cref{tab:denoised_results} the effectiveness of our LLM-based methods and the two upper bound baselines only \textbf{for the subset of users that were denoised}. In \cref{tab:denoised_results} we see that across all datasets, our LLM-based denoising approaches show greater improvements over the original user profiles, while still denoising a large portion of the users. On MovieLens 1M, denoising with Qwen few-shot-2 results in 6.7\% better HR@20 and up to 6.4\% better NDCG@20 for more than 50\% of the users. More interestingly, both Qwen few-shot-2 and Mistral few-shot-2 surpass even the upperBoundOnVal-2 baseline (recall that this is an upper bound baseline for the validation set item and not for the test set). On Amazon CDs \& Vinyl, Mistral zero-shot-recs-2 improves over the original user profiles by at most 13\% on HR@20 and up to 11.7\% for over 42\% of the users, bridging the gap with upperBoundOnVal-2 which still provides the best result. Similarly, on Yelp, despite the higher LLM hallucination rate even for users that were denoised, several LLMs manage to surpass the upperBoundOnVal baselines. These results clearly indicate that \textbf{our post-training denoising with LLMs can significantly improve pre-trained CF models without requiring any training or auxiliary data}.

\subsubsection{Effect of prompt formulation in post-training denoising (RQ2).} \label{sec:rq2}
\begin{figure}[htbp]
    \centering
    \includegraphics[width=0.95\linewidth]{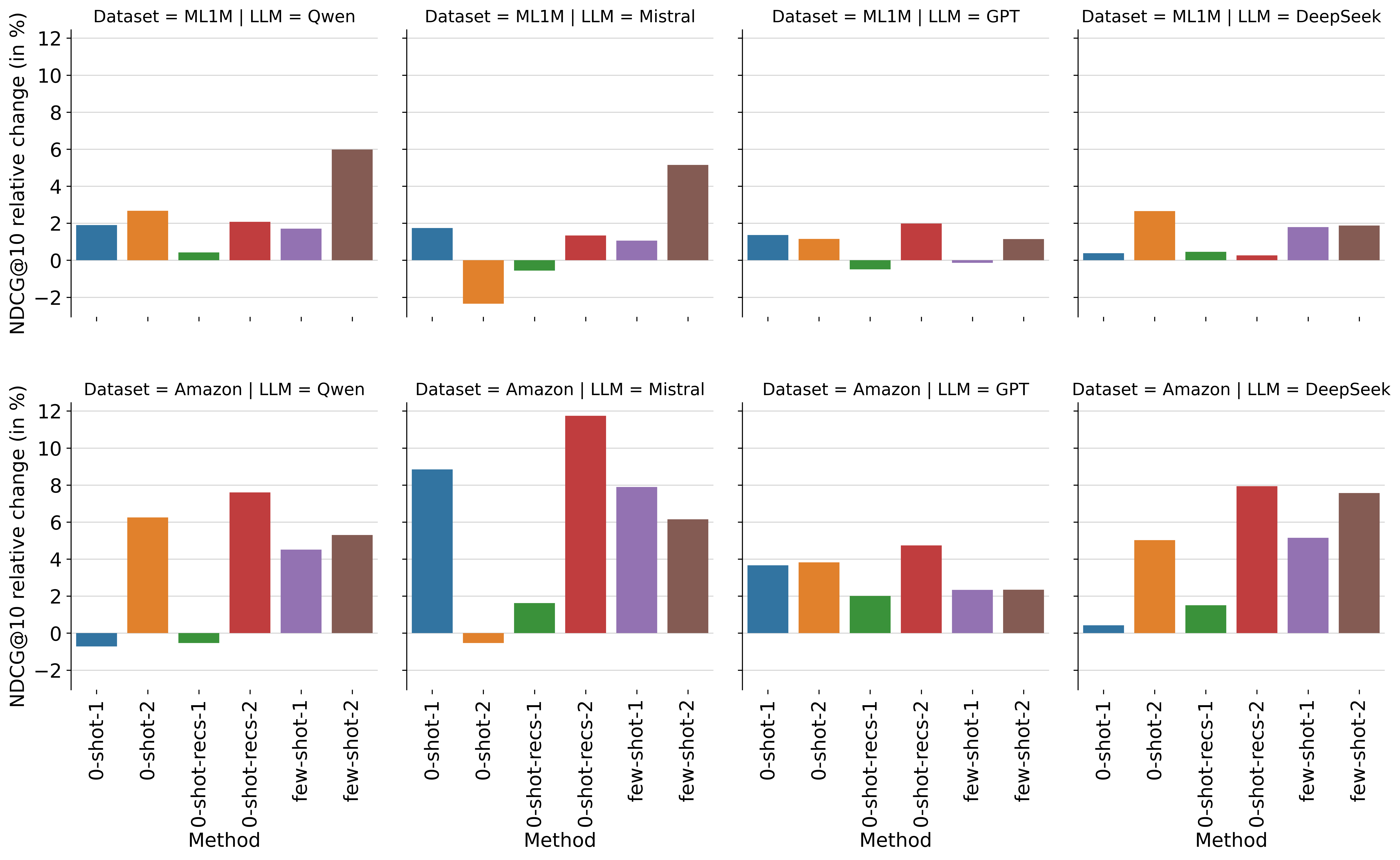}
    \caption{Relative change in NDCG@10 (in \%) for the different prompt formulations, on the MovieLens 1M and Amazon CDs \& Vinyl datasets,  for denoised user profiles.}
    \label{fig:prompt_bars}
\end{figure}

In \cref{fig:prompt_bars} we show the relative change (in \%) in NDCG@10 of denoised users across different prompt variations\footnote{For space, we show only MovieLens \& Amazon CDs \& Vinyl. Yelp has similar trends.}. We see that 
across datasets and LLMs, including either the top-10 recommendations by the CF model or denoising examples improves effectiveness; few-shot-2 and zero-shot-recs-2 yield the best results in the MovieLens 1M and Amazon CDs \& Vinyl datasets, respectively. For Qwen on Amazon CDs \& Vinyl, we see that zero-shot prompt has comparable performance to the prompts with examples, indicating that the rich parametric knowledge of the LLM provides enough information for the LLM to denoise the user profiles. Interestingly, GPT shows poor effectiveness on both datasets with little variation across prompt formulations. Qwen and Mistral provide larger improvements compared to the other LLMs on both datasets, with Mistral zero-shot-recs-2 yielding the highest relative change. Finally, we note that increasing the number of items removed from the user profiles from 1 to 2 results in larger improvements, with the exception of Mistral in the zero-shot scenario. Overall,~\cref{fig:prompt_bars} shows that, for our denoising task, \textbf{prompt optimality is LLM-specific and dataset-specific}, i.e. there is no single best prompting strategy for all LLMs and all datasets.

\subsubsection{Effect of item rating and user profile length on denoising (RQ3).} \label{sec:rq3}
\begin{figure}[htbp]
    \centering
    \includegraphics[width=0.85\linewidth]{figures/ml1m_denoised_ratings.png}
    \caption{Relative improvement (in \%) in NDCG@20 by ratings of removed items for denoised users in MovieLens 1M. Rating 5 is the best and 1 is the worst. Circle radius marks the number of users. Solid (resp. dashed) lines show positive (resp. negative) changes in NDCG@20 by denoising.}
    \label{fig:ml1m_denoised_ratings}

    \vspace{1em}
    \centering
    \includegraphics[width=0.85\linewidth]{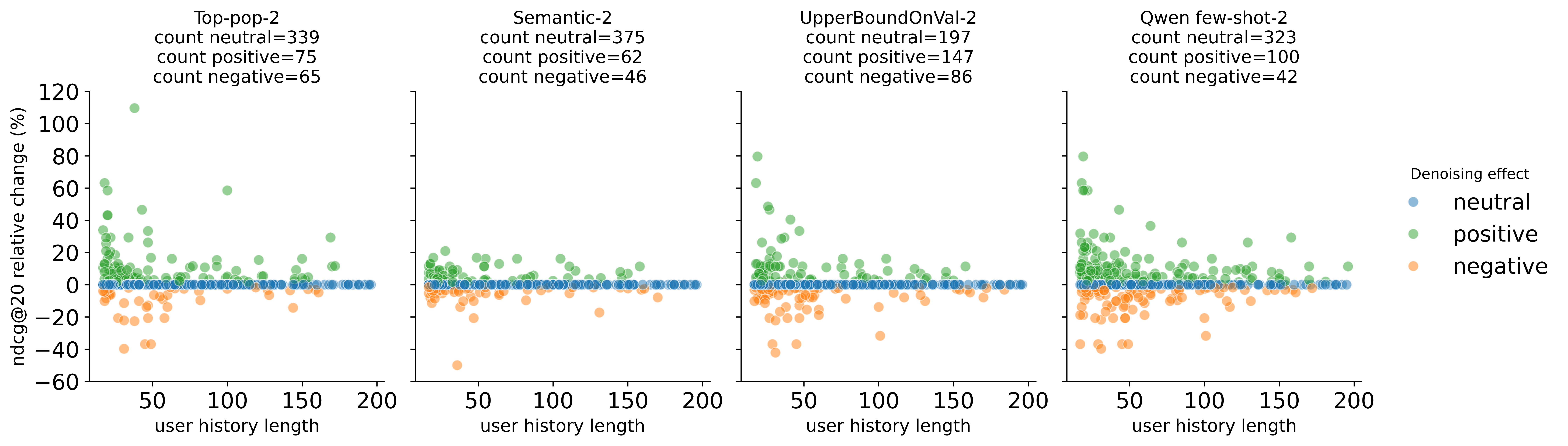}
    \caption{Relative improvement (in \%) in NDCG@20 by user profile length on denoised users in MovieLens 1M.}
    \label{fig:ml1m_denoised_history}
\end{figure}

In~\cref{fig:ml1m_denoised_ratings} we show the relative improvement in NDCG@20 when denoising with Qwen few-shot-2 and other baselines versus the item ratings of the removed items (by our denoising), for denoised users in MovieLens 1M\footnote{We include only MovieLens 1M for space reasons, other datasets follow similar trends overall.}. 
We see that the scale of improvement by our post-training denoising is much larger than the scale of deterioration; users benefit up to 15\% in relative improvement compared to up to -2.5\% negative change in NDCG@20 (these results are largely statistically significant, see~\cref{tab:denoised_results}). We also see that our approach can improve over original user profiles by removing items of different ratings, surpassing the upperBoundOnVal-2 baseline for items rated 3 and 5. This is also supported by~\cref{fig:ml1m_denoised_history}, where we show the relative change in NDCG@20 by the denoised user profile length, in MovieLens 1M. For each denoising method, we also give the number of users for which denoising improves, does not change, or negatively affects the rank of the candidate item. While few-shot-2 with Qwen improves fewer users than upper-bound-2 with respect to the candidate item, these users benefit from a larger gain in NDCG. More interestingly, our approach shows higher relative improvements on NDCG compared to upperBoundOnVal-2 for both users with short and longer profile lengths. Overall,~\cref{fig:ml1m_denoised_ratings,fig:ml1m_denoised_history} show that \textbf{our post-training denoising with LLMs benefits items of different ratings and users with different profile length}.

\section{Limitations and Conclusions}

In this work we provided the first post-training method for user profile denoising with LLMs in CF recommendations. Our approach results in up to 13\% improvement on effectiveness across 3 datasets whilst denoising up to 50\% of the users. Compared to previous in-training denoising, our method improves over original user profiles (no denoising) without retraining the CF model, fine-tuning the LLM, or requiring additional information. 

Our approach has some limitations. The candidate item used in the prompt to the LLMs can be itself a noisy interaction, which might make our approach less effective in denoising the remaining interactions in the user profile. In addition, due to the black box nature of LLMs, it is not straightforward to understand why the LLMs choose specific items to remove from the user interactions. For similar reasons, it is also not guaranteed that the LLM output will be in the correct format (as prompted) and will not include hallucinations. These are open problems in LLMs that also affect our work.  

Our work paves the way for post-training denoising approaches that make use of the parametric knowledge found in pre-trained LLMs, without requiring re-training the recommendation model. 

\bibliographystyle{splncs04}
\bibliography{bib}

@inproceedings{hu2008collaborative,
  title={Collaborative filtering for implicit feedback datasets},
  author={Hu, Yifan and Koren, Yehuda and Volinsky, Chris},
  booktitle={2008 Eighth IEEE international conference on data mining},
  pages={263--272},
  year={2008},
  organization={Ieee}
}

@inproceedings{joachims2017accurately,
  title={Accurately interpreting clickthrough data as implicit feedback},
  author={Joachims, Thorsten and Granka, Laura and Pan, Bing and Hembrooke, Helene and Gay, Geri},
  booktitle={Acm Sigir Forum},
  volume={51},
  number={1},
  pages={4--11},
  year={2017},
  organization={Acm New York, NY, USA}
}

@inproceedings{joachims2002optimizing,
  title={Optimizing search engines using clickthrough data},
  author={Joachims, Thorsten},
  booktitle={Proceedings of the eighth ACM SIGKDD international conference on Knowledge discovery and data mining},
  pages={133--142},
  year={2002}
}

@article{joachims2007evaluating,
  title={Evaluating the accuracy of implicit feedback from clicks and query reformulations in web search},
  author={Joachims, Thorsten and Granka, Laura and Pan, Bing and Hembrooke, Helene and Radlinski, Filip and Gay, Geri},
  journal={ACM Transactions on Information Systems (TOIS)},
  volume={25},
  number={2},
  pages={7--es},
  year={2007},
  publisher={ACM New York, NY, USA}
}

@inproceedings{wang2021denoising,
  title={Denoising implicit feedback for recommendation},
  author={Wang, Wenjie and Feng, Fuli and He, Xiangnan and Nie, Liqiang and Chua, Tat-Seng},
  booktitle={Proceedings of the 14th ACM international conference on web search and data mining},
  pages={373--381},
  year={2021}
}

@inproceedings{yi2014beyond,
  title={Beyond clicks: dwell time for personalization},
  author={Yi, Xing and Hong, Liangjie and Zhong, Erheng and Liu, Nanthan Nan and Rajan, Suju},
  booktitle={Proceedings of the 8th ACM Conference on Recommender systems},
  pages={113--120},
  year={2014}
}

@inproceedings{kim2014modeling,
  title={Modeling dwell time to predict click-level satisfaction},
  author={Kim, Youngho and Hassan, Ahmed and White, Ryen W and Zitouni, Imed},
  booktitle={Proceedings of the 7th ACM international conference on Web search and data mining},
  pages={193--202},
  year={2014}
}

@inproceedings{yang2012exploiting,
  title={Exploiting various implicit feedback for collaborative filtering},
  author={Yang, Byoungju and Lee, Sangkeun and Park, Sungchan and Lee, Sang-goo},
  booktitle={Proceedings of the 21st international conference on world wide web},
  pages={639--640},
  year={2012}
}

@inproceedings{liu2010understanding,
  title={Understanding web browsing behaviors through Weibull analysis of dwell time},
  author={Liu, Chao and White, Ryen W and Dumais, Susan},
  booktitle={Proceedings of the 33rd international ACM SIGIR conference on Research and development in information retrieval},
  pages={379--386},
  year={2010}
}

@inproceedings{jiang2020aspect,
  title={What aspect do you like: Multi-scale time-aware user interest modeling for micro-video recommendation},
  author={Jiang, Hao and Wang, Wenjie and Wei, Yinwei and Gao, Zan and Wang, Yinglong and Nie, Liqiang},
  booktitle={Proceedings of the 28th ACM International conference on Multimedia},
  pages={3487--3495},
  year={2020}
}

@inproceedings{wen2019leveraging,
  title={Leveraging post-click feedback for content recommendations},
  author={Wen, Hongyi and Yang, Longqi and Estrin, Deborah},
  booktitle={Proceedings of the 13th ACM Conference on Recommender Systems},
  pages={278--286},
  year={2019}
}

@article{han2018co,
  title={Co-teaching: Robust training of deep neural networks with extremely noisy labels},
  author={Han, Bo and Yao, Quanming and Yu, Xingrui and Niu, Gang and Xu, Miao and Hu, Weihua and Tsang, Ivor and Sugiyama, Masashi},
  journal={Advances in neural information processing systems},
  volume={31},
  year={2018}
}

@inproceedings{jiang2018mentornet,
  title={Mentornet: Learning data-driven curriculum for very deep neural networks on corrupted labels},
  author={Jiang, Lu and Zhou, Zhengyuan and Leung, Thomas and Li, Li-Jia and Fei-Fei, Li},
  booktitle={International conference on machine learning},
  pages={2304--2313},
  year={2018},
  organization={PMLR}
}

@inproceedings{gantner2012personalized,
  title={Personalized ranking for non-uniformly sampled items},
  author={Gantner, Zeno and Drumond, Lucas and Freudenthaler, Christoph and Schmidt-Thieme, Lars},
  booktitle={Proceedings of KDD cup 2011},
  pages={231--247},
  year={2012},
  organization={PMLR}
}

@inproceedings{kaplan2021unbiased,
  title={Unbiased filtering of accidental clicks in verizon media native advertising},
  author={Kaplan, Yohay and Krasne, Naama and Shtoff, Alex and Somekh, Oren},
  booktitle={Proceedings of the 30th ACM International Conference on Information \& Knowledge Management},
  pages={3878--3887},
  year={2021}
}

@inproceedings{wang2022learning,
  title={Learning robust recommenders through cross-model agreement},
  author={Wang, Yu and Xin, Xin and Meng, Zaiqiao and Jose, Joemon M and Feng, Fuli and He, Xiangnan},
  booktitle={Proceedings of the ACM web conference 2022},
  pages={2015--2025},
  year={2022}
}

@inproceedings{lin2023autodenoise,
  title={Autodenoise: Automatic data instance denoising for recommendations},
  author={Lin, Weilin and Zhao, Xiangyu and Wang, Yejing and Zhu, Yuanshao and Wang, Wanyu},
  booktitle={Proceedings of the ACM Web Conference 2023},
  pages={1003--1011},
  year={2023}
}

@article{song2024large,
  title={Large language model enhanced hard sample identification for denoising recommendation},
  author={Song, Tianrui and Chao, Wenshuo and Liu, Hao},
  journal={arXiv preprint arXiv:2409.10343},
  year={2024}
}

@inproceedings{wang2025unleashing,
  title={Unleashing the Power of Large Language Model for Denoising Recommendation},
  author={Wang, Shuyao and Zheng, Zhi and Sui, Yongduo and Xiong, Hui},
  booktitle={Proceedings of the ACM on Web Conference 2025},
  pages={252--263},
  year={2025}
}

@article{wang2025llm4dsr,
title = {LLM4DSR: Leveraging Large Language Model for Denoising Sequential Recommendation},
author = {Wang, Bohao and Liu, Feng and Zhang, Changwang and Chen, Jiawei and Wu, Yudi and Zhou, Sheng and Lou, Xingyu and Wang, Jun and Feng, Yan and Chen, Chun and Wang, Can},
year = {2025},
publisher = {Association for Computing Machinery},
address = {New York, NY, USA},
issn = {1046-8188},
journal = {ACM Trans. Inf. Syst.}
}

@inproceedings{hou2024large,
  title={Large language models are zero-shot rankers for recommender systems},
  author={Hou, Yupeng and Zhang, Junjie and Lin, Zihan and Lu, Hongyu and Xie, Ruobing and McAuley, Julian and Zhao, Wayne Xin},
  booktitle={European Conference on Information Retrieval},
  pages={364--381},
  year={2024},
  organization={Springer}
}

@inproceedings{deldjoo2024review,
  title={A review of modern recommender systems using generative models (gen-recsys)},
  author={Deldjoo, Yashar and He, Zhankui and McAuley, Julian and Korikov, Anton and Sanner, Scott and Ramisa, Arnau and Vidal, Ren{\'e} and Sathiamoorthy, Maheswaran and Kasirzadeh, Atoosa and Milano, Silvia},
  booktitle={Proceedings of the 30th ACM SIGKDD conference on Knowledge Discovery and Data Mining},
  pages={6448--6458},
  year={2024}
}

@article{zhao2024recommender,
  title={Recommender systems in the era of large language models (llms)},
  author={Zhao, Zihuai and Fan, Wenqi and Li, Jiatong and Liu, Yunqing and Mei, Xiaowei and Wang, Yiqi and Wen, Zhen and Wang, Fei and Zhao, Xiangyu and Tang, Jiliang and others},
  journal={IEEE Transactions on Knowledge and Data Engineering},
  volume={36},
  number={11},
  pages={6889--6907},
  year={2024},
  publisher={IEEE}
}

@article{wang2024towards,
  title={Towards next-generation llm-based recommender systems: A survey and beyond},
  author={Wang, Qi and Li, Jindong and Wang, Shiqi and Xing, Qianli and Niu, Runliang and Kong, He and Li, Rui and Long, Guodong and Chang, Yi and Zhang, Chengqi},
  journal={arXiv preprint arXiv:2410.19744},
  year={2024}
}

@inproceedings{bao2025customizing,
  title={Customizing In-context Learning for Dynamic Interest Adaption in LLM-based Recommendation},
  author={Bao, Keqin and Yan, Ming and Zhang, Yang and Zhang, Jizhi and Wang, Wenjie and Feng, Fuli and He, Xiangnan},
  booktitle={Findings of the Association for Computational Linguistics: ACL 2025},
  pages={14278--14291},
  year={2025}
}

@article{lin2025can,
  title={How can recommender systems benefit from large language models: A survey},
  author={Lin, Jianghao and Dai, Xinyi and Xi, Yunjia and Liu, Weiwen and Chen, Bo and Zhang, Hao and Liu, Yong and Wu, Chuhan and Li, Xiangyang and Zhu, Chenxu and others},
  journal={ACM Transactions on Information Systems},
  volume={43},
  number={2},
  pages={1--47},
  year={2025},
  publisher={ACM New York, NY}
}

@article{jarvelin2002cumulated,
  title={Cumulated gain-based evaluation of IR techniques},
  author={J{\"a}rvelin, Kalervo and Kek{\"a}l{\"a}inen, Jaana},
  journal={ACM Transactions on Information Systems (TOIS)},
  volume={20},
  number={4},
  pages={422--446},
  year={2002},
  publisher={ACM New York, NY, USA}
}

@inproceedings{ni2019justifying,
  title={Justifying recommendations using distantly-labeled reviews and fine-grained aspects},
  author={Ni, Jianmo and Li, Jiacheng and McAuley, Julian},
  booktitle={Proceedings of the 2019 conference on empirical methods in natural language processing and the 9th international joint conference on natural language processing (EMNLP-IJCNLP)},
  pages={188--197},
  year={2019}
}

@article{harper2015movielens,
  title={The movielens datasets: History and context},
  author={Harper, F Maxwell and Konstan, Joseph A},
  journal={Acm transactions on interactive intelligent systems (tiis)},
  volume={5},
  number={4},
  pages={1--19},
  year={2015},
  publisher={Acm New York, NY, USA}
}

@inproceedings{liang2018variational,
  title={Variational autoencoders for collaborative filtering},
  author={Liang, Dawen and Krishnan, Rahul G and Hoffman, Matthew D and Jebara, Tony},
  booktitle={Proceedings of the 2018 world wide web conference},
  pages={689--698},
  year={2018}
}

@article{kingma2013auto,
  title={Auto-encoding variational bayes},
  author={Kingma, Diederik P and Welling, Max},
  journal={arXiv preprint arXiv:1312.6114},
  year={2013}
}

@article{yang2025qwen3,
  title={Qwen3 technical report},
  author={Yang, An and Li, Anfeng and Yang, Baosong and Zhang, Beichen and Hui, Binyuan and Zheng, Bo and Yu, Bowen and Gao, Chang and Huang, Chengen and Lv, Chenxu and others},
  journal={arXiv preprint arXiv:2505.09388},
  year={2025}
}

@article{liu2024deepseek,
  title={Deepseek-v3 technical report},
  author={Liu, Aixin and Feng, Bei and Xue, Bing and Wang, Bingxuan and Wu, Bochao and Lu, Chengda and Zhao, Chenggang and Deng, Chengqi and Zhang, Chenyu and Ruan, Chong and others},
  journal={arXiv preprint arXiv:2412.19437},
  year={2024}
}

@inproceedings{dang2025data,
  title={Data augmentation as free lunch: Exploring the test-time augmentation for sequential recommendation},
  author={Dang, Yizhou and Liu, Yuting and Yang, Enneng and Huang, Minhan and Guo, Guibing and Zhao, Jianzhe and Wang, Xingwei},
  booktitle={Proceedings of the 48th International ACM SIGIR Conference on Research and Development in Information Retrieval},
  pages={1466--1475},
  year={2025}
}

@inproceedings{lin2023self,
  title={A self-correcting sequential recommender},
  author={Lin, Yujie and Wang, Chenyang and Chen, Zhumin and Ren, Zhaochun and Xin, Xin and Yan, Qiang and de Rijke, Maarten and Cheng, Xiuzhen and Ren, Pengjie},
  booktitle={Proceedings of the ACM Web Conference 2023},
  pages={1283--1293},
  year={2023}
}

\end{document}